\documentclass[
12pt,								
oneside,							
headsepline,					
appendixprefix,				
natbib 
]
{article}
\usepackage[english]{babel}
\usepackage{multicol}
\usepackage{graphicx} 	
\usepackage{amsmath}
 \usepackage{epstopdf} 
\usepackage{textcomp}
\usepackage{hyperref}
\usepackage{setspace}
\usepackage{lineno} 
\begin{document}

\title{Note on the classification of super-resolution in far-field microscopy and information theory}
\author{Oliver Passon\footnote{University of Wuppertal, School of Mathematics and Natural Science, Gau{\ss}str.~20, 42117 Wuppertal, passon@uni-wuppertal.de} and Johannes Grebe-Ellis\footnote{University of Wuppertal, School of Mathematics and Natural Science, Gau{\ss}str.~20, 42117 Wuppertal, grebe-ellis@uni-wuppertal.de}}

 \maketitle
 \begin{abstract}
 In recent years several far-field microscopy techniques have been developed which manage to overcome the diffraction limit of resolution. A unifying classification scheme for them is clearly desirable. We argue that existing schemes based on the information capacity of the optical system can not easily be extended to cover e.g., STED microscopy or techniques based on single molecule imaging. We suggest a classification based on a reconstruction of the Abbe limit.\footnote{Published version: J. Opt. Soc.  Am. A {\bf33}(7) B31-B35 \textcopyright\ 2016 Optical Society of America. One print or electronic copy may be made for personal use only. Systematic reproduction and distribution, duplication of any material in this paper for a fee or for commercial purposes, or modifications of the content of this paper are prohibited. \url{https://www.osapublishing.org/josaa/abstract.cfm?uri=josaa-33-7-B31}}
\end{abstract}

 \section{Introduction}
In 1873 Ernst Abbe established  his celebrated diffraction limit and showed that the resolution of an imaging system is limited by the wave-length $\lambda$ and the numerical aperture NA according to \cite{abbe73}: 
 \begin{eqnarray}
d_{\mathrm{Abbe}}=\frac{\lambda}{c\cdot {\mathrm{NA}}}, \label{equ_abbe}
\end{eqnarray}
with $c=1$ (coherent illumination) or $c=2$ (incoherent illumination).
Abbe's derivation was embeded in a model of the image formation which involves a double Fourier transformation \cite{steward2004,goodman96}. This is equivalent to an image formation described by a convolution of the object intensity (or amplitude), $O(x,y)$, with the PSF of the system (We restrict ourselves to the lateral plane):
\begin{eqnarray}
 I(x,y)  &=&   O(x,y) \otimes \mathrm{PSF}(x,y). \label{conv}
\end{eqnarray}
Applying a Fourier transformation and the convolution theorem yields:
\begin{eqnarray}
\tilde{I}(k_x,k_y) = \widetilde{O}(k_x,k_y)\cdot \mathrm{OTF}(k_x,k_y).      \label{otf}
\end{eqnarray}
Here, the image formation is described by the Fourier transform of the PSF, i.e., the optical transfer function (OTF). The Abbe resolution limit then amounts to the fact, that the cut-off of the optical transfer function (OTF) is given by $k_0=\frac{c\cdot NA}{\lambda}$. Super-resolution has become the collective term for techniques to break this limit and we will summarize some of these developments briefly in Sec.~2. Our focus will be put on applications in far-field microscopy. Given the diversity of different schemes a classification of them is desirable. In Sec.~3 we review briefly the existing suggestions based on information theory and point to their shortcomings in the current formulation. A novel classification scheme will be proposed in Sec.~4.

\section{Super-resolution imaging}
Already in the 1950s several researchers challenged this alleged fundamental Abbe limit. As pointed out by Wolter and Harris \cite{wolter58,wolter61,harris64b} the diffraction limit (Eq.~\ref{equ_abbe}) assumes an infinitely extended object. These authors noted, that the Fourier transform of a  {\em finite} function \cite[p.202]{wolter61} is analytic, i.e., given the {OTF} on a finite interval the transfer function can be recovered uniquely (and beyond the cut-off) by analytic continuation. This opens the possibility of super-resolution by deconvolution. However, noise renders the corresponding inverse problem ill-posed and practical applications of this ``computational super-resolution"  achieve only a minor resolution improvement \cite{bertero98,bertero2003}. But given that the finite extent of the probe is a generic property all this demonstrates that the resolution is only limited by noise -- contrary to Abbe's claim \cite{dekker97}.

Another early attempt to break the diffraction limit was made by Giuliano Toraldo di Francia. In \cite{toraldodifrancia52} it is shown that the width of the point-spread function can be reduced arbitrarily {\em without} increasing the side-lobs in the field-of-view. The idea is to apply a filtering or masking technique (called apodization). A further elaboration was given by Frieden \cite{frieden69}. However, the resulting central peak is very weak which limits the practical application of the idea \cite[p.448f]{lipson}. A recent implementation of this scheme is developed in \cite{huang09}.

Even in confocal microscopy (developed in the 1950s by the  cognitive scientist Marvin Minsky and later re-discovered, compare \cite{minsky}) the band-width of the OTF is  extended by a factor of 2, given that its effective PSF is the square of the illumination- and detection-PSF \cite{wilson2011}. Again, noise and the finite pin-hole size render the actual resolution gain
smaller or even vanishing. Confocal microscopy is a scanning technique, i.e., applies a non-uniform illumination of the sample. The idea to enhance resolution by non-uniform illumination (often called ``structured illumination") was suggested already in 1952 by the french physicist Maurice Fran\c{c}on. Some simple applications are discussed in \cite[p.472]{wolter58} and \cite{lukosz63} contains further developments of the concept (still restricted to coherent illumination). The idea of structured illumination microscopy is to apply an illumination which contains a spatial frequency $k_1$ and gives rise to the Moir\'{e} effect with fringes of frequency $|k-k_1|$ (with $k$ a sample frequency).   For $|k-k_1|<k_0$ these fringes will be observable in the miscroscope, i.e., effectively the passband is extended by $k_1$. Given that the highest frequency in the illumination pattern is as much diffraction limited as the detection passband the maximum value is $k_1=k_0$, hence the resolution can be extended by a factor of 2 (as in confocal microscopy; at least theoretically). However, a resolution beyond the Abbe limit needs incoherent illumination and the first discussion in the context of fluorescence microscopy (i.e., incoherent illumination) was given by Heintzmann and Cremer \cite{heintzmannCremer99}. The successful implementation and measurement results are reported in \cite{gustafsson2000}.

Structured illumination microscopy (SIM) can be improved further if the effects of non-linearity between excitation and emission are exploited \cite{hjc2002,heintzmannG2009}. As in linear SIM, higher frequency contributions can be moved into the passband of the original system. If the non-linearity is  non-polynomial the passband is (theoretically) even unbounded. Heintzmann et al. \cite{hjc2002} suggest to use the saturation of the fluorophore excitation as such a non-linearity (called SSIM for ``saturated structured illumination microscopy"). \cite{gustafsson2005} describes the practical implementation of this scheme and reports an experimental  resolution of $<$50nm (i.e., roughly a four-fold improvement compared to the Abbe limit). 

Fluorescence microscopy is also the arena for the latest developments. Two families can be distinguished which utilize the ability to switch fluorophores between different states (``on -- off" or ``bright -- dark"). Stimulated emission depletion (STED) microscopy excites the fluophores in a diffraction limited spot at first. However,  an additional STED beam de-excites all fluorophores but those in a small region close to the zero-point of the  doughnut-shaped STED beam  \cite{hell94,klar99}. This leads to a reduced  area of potential emittance. The width of the effective PSF depends on the intensity of the STED beam, $I_{\mathrm{max}}$, and the saturation intensity of the corresponding fluophores, $I_{\mathrm{sat}}$, according to \cite{harke08}:
\begin{eqnarray}
d_{\mathrm{STED}}=\frac{d_{\mathrm{Abbe}}}{\sqrt{1+\frac{I_{\mathrm{max}}}{I_{\mathrm{sat}}}}}. \label{equ_abbe_sted}
\end{eqnarray}
Stimulated emission is just one process to distinguish markers and the generalized class of microscopy techniques which exploit similar effects has been labeled RESOLFT (reversible saturable optical fluorescence transitions), \cite{hell2009}.

Finally, we should mention the family of microscopy techniques based on single molecule imaging (SMI). As suggested by Betzig the basic idea is the separation of nearby fluorophores  through ``unique optical characteristics" \cite[p.237]{betzig95}.  To achieve this separation through the time domain was accomplished for the first time by  Lidke et al. \cite{lidkeetal2005} through the ``blinking" of quantum dots. For biological imaging the successful application was reported in 2006 independently by three groups. Their methods were named STORM (Stochastic Optical Reconstruction Microscopy) \cite{rust2006}, PALM (Photo Activated localization Microscopy) \cite{betzig2006} and FPALM (Fluorescence Photo Activated localization Microscopy) \cite{hess2006}. In the mean time other modified schemes of localization microscopy have been developed like dSTORM (direct STORM) or PALMIRA (PALM with independently running acquisition). 

All of them apply a similar strategy: The probe is labeled with photo-switchable fluorescent markers and a weak light pulse activates a random, sparse subset of these fluorophores. Ideally each of these sources is separated by more than the Abbe limit. However,  for an emitter known to be isolated the {\em localization} precision is not restricted by diffraction. Given the shape of the PSF its mean can be estimated from a fit to the data with a precision limited by the signal intensity and SNR only. For the detection of a full image  a strong ``bleaching" pulse is applied to make the active molecules permanently (or temporary) dark. Another activation pulse then turns on a {\em different} sparse subset, which is again localized. This cycle is repeated until sufficient image details have been acquired or all  the dye molecules have been switched. This results into a list of emitter positions, localization precisions, noise and background. This data can be used to render an image which shows details  with a resolution between 10 and 50nm \cite[p.201]{thorley2014}. 

\section{Classification based on information theory}
Our brief summary of far-field microscopy techniques to break the Abbe limit illustrated the diversity of approaches. However, it has been noted by Testorf and Fiddy \cite[p.166f]{testorf2010} that while most of them have been labeled as ``super-resolution", this has been done 
\begin{quote}
``
[...] frequently independent of adherence to any of the conditions that would define a meaningful resolution limit in classical terms. [...] 
This, in turn, has created a culture of reporting on new superresolution schemes in terms of the achievable resolution rather than in terms of the relationships to preexisting methods or to fundamental underlying assumptions."
\end{quote}
They conclude, that the comparison of different schemes is hindered and that a  proper and systematic classification of existing methods would even aid the development of new methods.  In a similar vein Sheppard \cite{sheppard07} has noted that the concept of super-resolution is ``somewhat confused". 

In \cite{testorf2010} and  \cite{sheppard07} a classification based on information theory is applied to arrive at a unified framework for the discussion of super-resolution and we will briefly review this work. According to Cox and Sheppard \cite{cox86} the information capacity of an optical system is given by:
\begin{eqnarray}
N=(2L_x B_x+1)(2L_y B_y+1)(2L_z B_z+1) (2TB_T+1) \cdot  \log (1+\mathrm{SNR}) \label{ic}
\end{eqnarray}
Here, $B_x$, $B_y$ and $B_z$ denote the spatial band-width in the corresponding direction, $B_T$ the temporal band-width, $T$ the observation time, $L_xL_y$ the field-of-view and $L_z$ the depth-of-field of the system. SNR denotes the signal-to-noise ratio. While this information capacity can not be exceeded, which is called the ``Theorem of Invariance of Information Capacity" \cite[p.1154]{cox86}, the spatial band-width can be increased at the expense of e.g., the temporal band-width or the field-of-view. In \cite{testorf2010,cox86,sheppard07} this approach is used to make the trade-offs transparent which underly different super-resolution schemes. 

The super-resolving pupils suggested in \cite{toraldodifrancia52} increase the resolution at the expense of the reduced field-of-view since they produce side-lobes. Thus, in themselves (i.e., without restricting the field-of-view) these pupils do not increase the spatial frequency band-width, i.e., they are not super-resolving {\em per se}. This is why Cox and Sheppard \cite{cox86} suggest the term ``ultraresolution" for them. In contrast, do confocal microscopy or SIM (briefly touched uppon in Sec.~2) increase the spatial frequency band-width at the expense of the temporal band-width, i.e., both methods need several images to be taken (confocal microscopy is a scanning technique and SIM needs several images with rotated geometry of the structured illumination). Note, that the latter methods assume some prior knowldege about the object, namely that it does not vary in time. Only then the temporal band-width can be traded against the spatial band-width. We note in passing that also confocal or structured illumination microscopy are not super-resolving {\em per se} until the additional images have been recorded. Why a similar argument qualifies Toraldo di Francia's pupils an instance of ``ultraresolution" only (i.e., no super-resolution proper) appears questionabe to us. 

Cox and Sheppard \cite{cox86} apply the information capacity approach also in the case of unrestricted super-resolution provided by analytic continuation. Here, the trade-off is with respect to the SNR, given that the spectrum within the passband is not know with arbitrary precision and the limit is set by noise.

Summing up, Sheppard \cite{sheppard07} suggests three types of super-resolution: (i) improved spatial frequency response with unchanged  cut-off (as with superresolving pupils introduced by Toraldo di Francia) (ii) techniques with increased cut-off (like e.g., confocal scanning laser microscopy and SIM) and (iii) unrestricted super-resolution by fundamentally unconfined  increase of the cut-off of the OTF (e.g., by analytic continuation).

However, how do the more recent techniques like STED, SSIM or single molecule imaging (SMI)  fit into this picture?  While STED and SSIM are briefly mentioned in \cite{sheppard07} this work was apparently written before the advent of STORM, PALM and FPALM. SSIM and STED may be related to this classification by noting that also here the {\em unrestricted} resolution needs a trade-off with the SNR. Given that  the photo-chemical and spectral properties of the fluorophores provide a piece of information about the object they apparently fit well into the conceptual framework of information theory. This holds at least if one views the fluorophores as the ``object" of fluorescence microscopy -- and not the structure which has been labeled. Otherwise the fluorophores are actually part of the imaging system. However, how to quantify this information in the current framework is rather unclear and one may conjecture that an additional term needs to be included into  Eq.~\ref{ic}. Key to the novel superresolving techniques in fluorescence microscopy is to exploit the (non-linear) {\em interaction} with the contrast generating agent. One may also argue that the concept of an imaging system as a (passive) information channel is not sufficient to capture this novel aspect of imaging. 

Be this as it may, we will leave this question aside and turn to single molecule imaging (SMI) which is more complicated still. Note, that the discussion so far was framed in terms of isoplanatic systems in which the PSF does not depend on the position. Only then the image formation can be described by a convolution (Eq.~\ref{conv}). However, this assumption does not apply in 
single molecule imaging. In the first place, these methods produce initially no image at all but a data set of emitter positions, localization  precisions and  intensities (signal and background). Agreed, all image-data acquisition systems produce ``data" in the first place, but here  there is no natural way to display this data and they need to be rendered according to some user specified method (See e.g., \cite[p.418ff]{nsr14} for a discussion of several visualization methods and their interrelation to the resolution issue). 
A natural candidate for the effective PSF in SMI microscopy is apparently a point-spread function (e.g., in the Gaussian approximation) with the  localization precision as its width. However, this  ignores  that in general SMI microscopy yields a different precision for each fitted emitter. Thus, the image formation can not be described by a convolution with a PSF, since there is no {\em single} (effective) PSF to convolve with. Only if one {\em neglects} the position dependence of the effective PSF, SMI can be described by a convolution (see e.g., \cite[p.16]{middendorff2008} where PALMIRA is treated). However, the quality of this approximation needs to be tested case by case. 
Hence, the strategy of SMI microscopy to break the Abbe limit is not covered by the classification suggested in \cite{sheppard07}.

One may wonder how the resolution in SMI microscopy is defined, given that the common approach via the OTF cut-off is blocked. In fact, usually the mean localization precision (e.g., in terms of the full width at half maximum \cite{rust2006}), the resolution derived from the labeling density by the Nyquist criterion \cite{shroff2008} or a combination of both \cite{laka2012} are quoted. However, all these definitions are heuristic only. Given these deep conceptual problems other integral resolution measures have been suggested.  For example \cite{rees2012,fitzgerald2012} base their different approaches on estimation theory while in \cite{banterle2013,netal13} the Fourier ring correlation (FRC) is proposed as a resolution measure for SMI microscopy.

\section{Novel classification by reconsidering the Abbe limit} 
In order to classify the strategies for breaking the Abbe limit properly we therefore suggest to spell out its content more carefully.  It is certainly well known, that in its current Fourier optical formulation the Abbe limit can be construed as a conjunction of three hierarchically related claims: 
\begin{quote} 
{\bf Convolution assumption (CONV)},\\ 
whereby the  image formation can be represented as a convolution of the object distribution with a point-spread function (PSF). 
 
{\bf Resolution-cut-off relation (RCR)},\\ 
whereby the principle resolution limit is given by the cut-off frequency of the Fourier transform of the PSF (i.e., the OTF).
 
{\bf Abbe Cut-off  (AC)},\\ whereby the cut-off frequency of an optical system is given by $k_0={\mathrm{NA}}/\lambda$ (coherent case) or  $k_0=2{\mathrm{NA}}/\lambda$ (incoherent case).
 \end{quote}
Now it becomes evident that ``breaking the Abbe limit" can mean quite different things. E. g. confocal microscopy, linear structured illumination, but also STED microscopy or SSIM expand the band-width explicitly or effectively. Here the claim AC is refuted while RCR (and also CONV) remain the underlying assumption and motivation. One might describe the situation by saying that this breaking or ``bypassing" of the Abbe limit is still guided by Abbe's original reasoning. 
 
In contrast, the observation that by analytic continuation the OTF can (in principle) be extrapolated  {\em beyond} the cut-off frequency refutes the claim RCR. Note, that the cut-off is not altered by this procedure and that the actual passband provided by any specific optical system (e.g., according to AC) plays no role whatsoever.  

Finally, in SMI microscopy (e.g., STORM, PALM or FPALM) the underlying convolution assumption (CONV) regarding the image formation does not apply. Expressed pointedly, if there is no OTF, it can neither be used to define the resolution nor can its  cut-off  be extended.

\section{Summary and conclusion} 
We have argued that it is difficult to apply the current information theoretic framework to deal with the recent developments in super-resolving fluorescense microscopy. It remains true that all kinds of super-resolution exploit some prior knowledge about the sample but to incorporate this into the information theoretic framework needs to be an objective of future developments in this field.  E.g., STED microscopy utilizes properties of the fluorophores which can not be quantified in the current information theoretical formalism.    In SMI microscopy the very definition of the band-width of the imaging system is intricate. Thus, also here does the current information theoretic framework not help to make the corresponding trade-offs transparent. Our simple reconstruction of the Abbe limit as the conjunction of the (i) convolution assumption, (ii) the OTF-based resolution definition and (iii) the specific OTF cut-off values as derived by Abbe allows for a more differentiated description of the strategies to break this limit. This perspective may provide a complementary view on the classification in super-resolution microscopy. 

\subsection*{Acknowledgement}
We gratefully acknowledge the illuminating email exchange with  Marcel Lauterbach, discussions with Marc M\"uller and helpfull comments by two anonymous referees.


\begin{thebibliography}{10}
\newcommand{\enquote}[1]{``#1''}
\begin{singlespace}
\bibitem{abbe73}
E.~Abbe, \enquote{Beitr\"age zur Theorie des Mikroskops und der mikroskopischen Wahrnehmung,} Archiv f\"ur mikroskopische Anatomie \textbf{9}, 413--468 (1873).
\bibitem{steward2004}
E.~G. Steward, \emph{Fourier Optics -- An Introduction} (Dover, 2004).
\bibitem{goodman96}
J.~W. Goodman, \emph{Introduction to Fourier Optics} (McGraw-Hill, 1996).
\bibitem{wolter58}
H.~Wolter, \enquote{Zum grundtheorem der imformationstheorie, insbesondere in
  der optik,} Physica \textbf{24}, 457--475 (1958).
\bibitem{wolter61}
H.~Wolter, \enquote{On basic analogies and principal differences between
  optical and electronic information,} in \enquote{Progress in Optics,} ,
  E.~Wolf, ed. (North-Holland, Amsterdam, 1961), chap.~5, pp. 157--210.

\bibitem{harris64b}
J.~L. Harris, \enquote{Diffraction and resolving power,} J. Opt. Soc. Am.
  \textbf{54}, 931--933 (1964).

\bibitem{bertero98}
M.~Bertero and P.~Boccacci, \emph{Introduction to Inverse Problems in Imaging}
  (Institute of Physics Publishing, 1998).

\bibitem{bertero2003}
M.~Bertero and P.~Boccacci, \enquote{Super-resolution in computational
  imaging,} Micron \textbf{34}, 265--273 (2003).

\bibitem{dekker97}
A.~J. den Dekker and A.~van~den Bos, \enquote{Resolution: A survey,} J. Opt.
  Soc. Am. A \textbf{14}, 547--557 (1997).

\bibitem{toraldodifrancia52}
G.~T. di~Francia, \enquote{Super-gain antennas and optical resolving power,}
  Suppl. Nuovo Cimento \textbf{9}, 426--438 (1952).

\bibitem{frieden69}
B.~R. Frieden, \enquote{On arbitrarily perfect imagery with a finite aperture,}
  Opt. Act. \textbf{16}, 795--807 (1969).

\bibitem{lipson}
A.~Lipson, S.~G. Lipson, and H.~Lipson, \emph{Optical Physics} (Cambridge
  University Press, 2010).

\bibitem{huang09}
F.~M. Huang and N.~I. Zheludev, \enquote{Super-resolution without evanescent
  waves,} Nano Lett. \textbf{9}, 1249--1254 (2009).

\bibitem{minsky}
M.~Minsky, \enquote{Memoir on inventing the confocal scanning microscope,}
  Scanning \textbf{10}, 128--138 (1988).

\bibitem{wilson2011}
T.~Wilson, \enquote{Resolution and optical sectioning in the confocal
  microscope,} Journal of Microscopy \textbf{244}, 113--121 (2011).

\bibitem{lukosz63}
W.~Lukosz and M.~Marchand, \enquote{Optischen abbildung unter \"uberschreitung
  der beugungsbedingten aufl\"osungsgrenze,} Opt. Act. \textbf{10}, 241--255
  (1963).

\bibitem{heintzmannCremer99}
R.~Heintzmann and C.~Cremer, \enquote{Laterally modulated excitation
  microscopy: improvement of resolution by using a diffraction grating,} Proc.
  SPIE \textbf{3568}, 185--196 (1999).

\bibitem{gustafsson2000}
M.~G.~L. Gustafsson, \enquote{Surpassing the lateral resolution limit by a
  factor of two using structured illumination microscopy,} Journal of
  Microscopy \textbf{198}, 82--87 (2000).

\bibitem{hjc2002}
R.~Heintzmann, T.~M. Jovin, and C.~Cremer, \enquote{Saturated patterned
  excitation microscopy -- a concept for optical resolution improvement,} J.
  Opt. Soc. Am. A \textbf{19}, 1599--1609 (2002).

\bibitem{heintzmannG2009}
R.~Heintzmann and M.~G.~L. Gustafsson, \enquote{Subdiffraction resolution in
  continuous samples,} Nature Photonics \textbf{3}, 362--364 (2009).

\bibitem{gustafsson2005}
M.~G.~L. Gustafsson, \enquote{Nonlinear structured-illumination microscopy:
  wide-field fluorescence imaging with theoretically unlimited resolution,}
  Proc. Nat. Acad. Sci. USA \textbf{102}, 13081--13086 (2005).

\bibitem{hell94}
S.~W. Hell and J.~Wichmann, \enquote{Breaking the diffraction resolution limit
  by stimulated emission: stimulated-emission-depletion fluorescence
  microscopy,} Optics Letters \textbf{19}, 780--782 (1994).

\bibitem{klar99}
T.~A. Klar and S.~W. Hell, \enquote{Subdiffraction resolution in far-field
  fluorescence microscopy,} Optics Lett. \textbf{24}, 954--956 (1999).

\bibitem{harke08}
B.~Harke, J.~Keller, C.~K. Ullal, V.~Westphal, A.~Sch\"onle, and S.~W. Hell,
  \enquote{Resolution scaling in sted microscopy,} Optical Express \textbf{16},
  4154--4162 (2008).

\bibitem{hell2009}
S.~W. Hell, \enquote{Microscopy and its focal switch,} Nature Methods
  \textbf{6}, 24--32 (2009).

\bibitem{betzig95}
E.~Betzig, \enquote{Proposed method for molecular optical imaging,} Optics
  Letters \textbf{20}, 237--239 (1995).

\bibitem{lidkeetal2005}
K.~A. Lidke, B.~Rieger, T.~M. Jovin, and R.~Heintzmann,
  \enquote{Superresolution by localization of quantum dots using blinking
  statistics,} Optics Express \textbf{13}, 7052--7062 (2005).

\bibitem{rust2006}
M.~J. Rust, M.~Bates, and X.~Zhuang, \enquote{Sub-diffraction-limit imaging by
  stochastic optical reconstruction microscopy (storm),} Nature Methods
  \textbf{3}, 793--796 (2006).

\bibitem{betzig2006}
E.~Betzig, G.~H. Patterson, R.~Sougrat, O.~W. Lindwasser, S.~Olenych, J.~S.
  Bonifacino, M.~W. Davidson, J.~Lippincott-Schwartz, and H.~F. Hess,
  \enquote{Imaging intracellular fluorescent proteins at nanometer resolution,}
  Science \textbf{313}, 1642--1645 (2006).

\bibitem{hess2006}
S.~T. Hess, T.~P.~K. Girirajan, and M.~D. Mason, \enquote{Ultra-high resolution
  imaging by fluorescence photoactivation localization microscopy,} Biophys J.
  \textbf{91}, 4258--4272 (2006).

\bibitem{thorley2014}
J.~A. Thorley, J.~Pike, and J.~Z. Rappoport, \enquote{Super-resolution
  microscopy: A comparison of commercially available options,} in
  \enquote{Fluorescence Microscopy -- Super-Resolution and other novel
  Techniques,}  A.~Cornea and P.~M. Conn, eds. (Elsevier, Amsterdam, 2014),
  chap.~14, pp. 199--212.

\bibitem{testorf2010}
M.~E. Testorf and M.~A. Fiddy, \enquote{Superresolution imaging -- revisited,}
  in \enquote{Advances in Imaging and Electron Physics Vol. 163,}  P.~W.
  Hawkes, ed. (Elsevier, Amsterdam, 2010), chap.~5, pp. 165--218.

\bibitem{sheppard07}
C.~J.~R. Sheppard, \enquote{Fundamentals of superresolution,} Micron
  \textbf{38}, 165--169 (2007).

\bibitem{cox86}
I.~J. Cox and C.~J.~R. Sheppard, \enquote{Information capacity and resolution
  in an optical system,} J. Opt. Soc. Am. A \textbf{3}, 1152--1158 (1986).

\bibitem{nsr14}
R.~P.~J. Nieuwenhuizen, S.~Stallinga, and B.~Rieger, \enquote{Visualization and
  resolution in localization microscopy,} in \enquote{Cell Membrane
  Nanodomains: from Biochemistry to Nanoscopy,} , A.~Cambi and D.~S. Lidke,
  eds. (Taylor and Francis, Boca Raton, 2014), chap.~18, pp. 409--430.

\bibitem{middendorff2008}
C.~von Middendorf, \enquote{Experimental stochastics in high-resolution
  fluorescence microscopy,} Ph.D. thesis, Uni Heidelberg,
  http://www.ub.uni-heidelberg.de/archiv/8736 (2008).

\bibitem{shroff2008}
H.~Shroff, C.~G. Galbraith, J.~A. Galbraith, and E.~Betzig, \enquote{Live-cell
  photoactivated localization microscopy of nanoscale adhesion dynamics,}
  Nature Methods \textbf{5}, 417--423 (2008).

\bibitem{laka2012}
M.~Lakadamyali, H.~Babcock, M.~Bates, X.~Zhuang, and J.~Lichtman, \enquote{3d
  multicolor super-resolution imaging offers improved accuracy in neuron
  tracing,} PLoS ONE \textbf{7}, e30826 (2012).

\bibitem{rees2012}
E.~Rees, M.~Erdelyi, D.~Pinotsi, A.~Knight, D.~Metcalf, and C.~Kaminski,
  \enquote{Blind assessment of localisation microscope image resolution,}
  Optical Nanoscopy \textbf{1}, 12 (2012).

\bibitem{fitzgerald2012}
J.~E. Fitzgerald, J.~Lu, and M.~J. Schnitzer, \enquote{Estimation theoretic
  measure of resolution for stochastic localization microscopy,} Phys. Rev.
  Lett. \textbf{109}, 048102 (2012).

\bibitem{banterle2013}
N.~Banterle, K.~H. Bui, E.~A. Lemke, and M.~Beck, \enquote{Fourier ring
  correlation as a resolution criterion for super-resolution microscopy,}
  Journal of Structural Biology \textbf{183}, 363--367 (2013).

\bibitem{netal13}
R.~P.~J. Nieuwenhuizen, K.~Lidke, M.~Bates, D.~L. Puig, D.~Gr\"unwald,
  S.~Stallinga, and B.~Rieger, \enquote{Measuring image resolution in optical
  nanoscopy,} Nature Methods \textbf{10}, 557--562 (2013).
\end{singlespace}
\end{thebibliography}
 \end{document}